# MAX Phase Zr$_2$SeC and Its Thermal Conduction Behavior

*Ke Chen [a,d,\*,#], Xiaojing Bai [b,#], Xulin Mu [c], Pengfei Yan [c], Nianxiang Qiu [a,d], Youbing Li [a,d], Jie Zhou [a], Yujie Song [a,d], Yiming Zhang [a,d], Shiyu Du [a,d], Zhifang Chai [a,d], Qing Huang [a,d,\*]*

[a] Engineering Laboratory of Advanced Energy Materials, Ningbo Institute of Materials Technology and Engineering, Chinese Academy of Sciences, Ningbo 315201, China.

[b] School of Materials Science and Engineering, Anyang Institute of Technology, Anyang 455000, China.

[c] Beijing Key Laboratory of Microstructure and Properties of Solids, Beijing University of Technology, Beijing 100124, China.

[d] Qianwan Institute of CNiTECH, Ningbo 315336, China.

[#] These authors contributed equally to this work.

[*] Corresponding authors.

E-mail addresses: chenke@nimte.ac.cn (Ke Chen), huangqing@nimte.ac.cn (Qing Huang).

**Abstract**

The elemental diversity is crucial to screen out ternary MAX phases with outstanding properties via tuning of bonding types and strength between constitutive atoms. As a matter of fact, the interactions between M and A atoms largely determine the physical and chemical properties of MAX phases. Herein, Se element was experimentally realized to occupy the A site of a MAX phase, Zr$_2$SeC, becoming a new member within this nanolaminated ternary carbide family. Comprehensive characterizations including Rietveld refinement of X-ray Diffraction and atom-resolved transmission electron microscopy techniques were employed to validate this novel MAX phase. The distinct thermal conduction behaviors emerged are attributed to the characteristic interactions between Zr and Se atoms.

**Keywords**: MAX phase, Zr$_2$SeC, Electrical property, Thermal property

## 1. Introduction

MAX phases are a large family of damage-tolerant structural materials, which have attracted great attentions in promising applications such as pantograph component and braking discs for high-speed trains [1], accident tolerant fuel cladding [2], and high-temperature electrodes. [3] These layered ternary compounds have a general chemical formula $M_{n+1}AX_n$; wherein M is an early transition metal, A is normally an A-group element, X is carbon and/or N, and n equals to 1-3. [1, 4] Due to diversity in composing elements and their free combinations, more than 150 MAX phases so far have been discovered. [1, 5] Besides the edge-sharing octahedrons [$M_6X$], the bonding between M- and A- elements plays an important role in determining physical and chemical properties of MAX phases. For instance, the abundant *d* or *s* electrons of A-site element might realize unexplored functions of MAX phases. H. Fashandi *et al.* [3] reported that high-temperature-stable Ohmic contacts between SiC and MAX phases were largely improved when noble-metal elements Au and Ir substituted Si in $Ti_3SiC_2$. Our previous reports proved a facile substitution of Zn and Cu for Al in traditional MAX phases using $ZnCl_2$ and $CuCl_2$ Lewis acid molten salts. [6-8] Furthermore, the A-site occupancy of Cu atoms in the $Ti_3(Al_xCu_{1-x})C_2$ showed compelling peroxidase-like catalytic activity for decomposing $H_2O_2$ into $O_2$. [6]

MAX phases with chalcogen elements at A site are very unique due to their strong chemical bondings between the outermost *p* electrons of chalcogen elements and the outermost *d* electrons of M-site atoms, where in common MAX phases (e.g. $Ti_2AlC$) are relatively weak. [9] Accordingly, $Ti_2SC$ shows higher Young's modulus and shear modulus, as well as enhanced anti-corrosive behavior (loss of A element) than $Ti_2AlC$ phase. [10-12] The chalcogen-containing ternary MAX phases thus far are limited to $Ti_2SC$, $Zr_2SC$, $Hf_2SC$, and $Nb_2SC$ [1, 13-15]. In this work, a novel chalcogen-containing ternary MAX phase, $Zr_2SeC$, was experimentally synthesized, expanding the options of chalcogen elements for MAX phases. The electrical and thermal properties of $Zr_2SeC$ phase were measured and discussed to show the role of A-site elements on the physical properties of MAX phase.

## 2. Experimental Details

2.1 Synthesis details

The $Zr_2SeC$ bulk was synthesized using commercially available powders. Considering the low boiling point of selenium, the zirconium diselenide was first synthesized using zirconium (99.5% purity, 400 mesh, Targets Research Center of General Research Institute for Nonferrous Metals, China) and selenium (99.99% purity, 200 mesh, Aladdin) powders in a sealed quartz tube at 900 °C (The details could be referred to previous reference. [16]) Then, the as-synthesized zirconium diselenide powders were mixed with zirconium and carbon (99.9% purity, 400 nm, Macklin, China) powders in an agate mortar with the molar ratio of 1.05:3:1.95 for 30 min under Ar atmosphere. The ground powder mixture was cold-pressed in a graphite mold. The $Zr_2SeC$ bulk was *in situ* sintered in a Pulse-Electric-Current-Aided sintering device (HP D 25/3, FCT Group, Germany) under Ar atmosphere. The initial heating rate applied on these green pellets from 450 °C to 1100 °C was 50 °C/min, and the rate following from 1100 °C to 1500 °C was 25 °C/min. The pressure (48 MPa) was loaded on the pellets along with the temperature rising, and kept at target temperature for 20 min.

A fully dense $Zr_2SC$ bulk was also sintered for a comparative analysis. The $Zr_2SC$ phase powder was synthesized using commercially ferrous sulfide (99% purity, 200 mesh, J&K Scientific Ltd, China), zirconium, and carbon powders with the molar ratio of 1.05:2:0.95. The same details were applied for $Zr_2SeC$ phase. The as-synthesized pellets were further pulverized and immersed in hydrochloric acid solution (1 M) for 2 days at 45 °C to get rid of the iron particles. The iron-removed powders were used to prepare dense $Zr_2SC$ monolith in the Pulse-Electric-Current-Aided sintering device. The pellets were sintered at 1600 °C for 20 min under Ar atmosphere, and the loading pressure was 48 MPa.

## 2.2 Computational details

The calculations were performed using density functional theory (DFT) [17] and plane-wave pseudopotential method implemented in the CASTEP (Cambridge Serial Total Energy Package) codes [18]. The generalized gradient approximation (GGA) [19] of Perdew-Breke-Ernzerh (PBE) [20] was adopted for the exchange and correlation energy [21]. Norm-conserving pseudopotentials with a 520 eV plane-wave cutoff energy were employed to explicitly treat valence electrons for Zr $4d^25s^2$, S $3s^23p^4$, Se $4s^24p^4$, and C $2s^22p^2$. The equilibrium structures were obtained via geometry optimization in the Broyden-Fletcher-Goldfarb-Shanno (BFGS) minimization scheme. The change of total energy during the self-consistent field (SCF) calculation was converged to $1.0\times10^{-7}$ eV/atom. All of our models were relaxed until the Hellman-Feynman forces on each atom were smaller than 0.002 eV/Å, atomic displacements were smaller than $1\times10^{-3}$ Å, and the stresses are less than 0.05 GPa.

## 2.3 Materials Characterization

The as-prepared powders and bulks were characterized by an X-ray diffractometer (XRD; Bruker AXS D8 Advance, Germany) with Cu K$_\alpha$ radiation operating at 40 kV and 40 mA at a step scans of 0.02° 2θ and a step time of 0.2 sec. The Rietveld method was employed to refine the crystal parameters and phase composition using the software of *TOPAS-Academic v6*. The morphology, microstructure, and chemical composition were observed by a field emission scanning electron microscope (SEM; FEI QUANTA 250 FEG, USA) and a transmission electron microscope (TEM; FEI Talos and Titan, USA) equipped with energy dispersive spectroscopy (EDS). The samples observed in TEM was cut and thinned using a dual-beam scanning electron microscope focused ion beam (FIB; Thermo scientific Helios-G4-CX, USA). Archimedes method in ambient temperature ethanol was employed to measure the density of as-prepared monoliths. The thermal properties were determined by laser flash technique (NETZSCH LFA467, Germany) from room temperature to 600 K (a standard graphite was used as reference for heat capacity). The electrical resistivity was detected by Physical Property Measurement System (PPMS; Quantum Design, USA) using a four-probe method.

## 3. Results and discussion

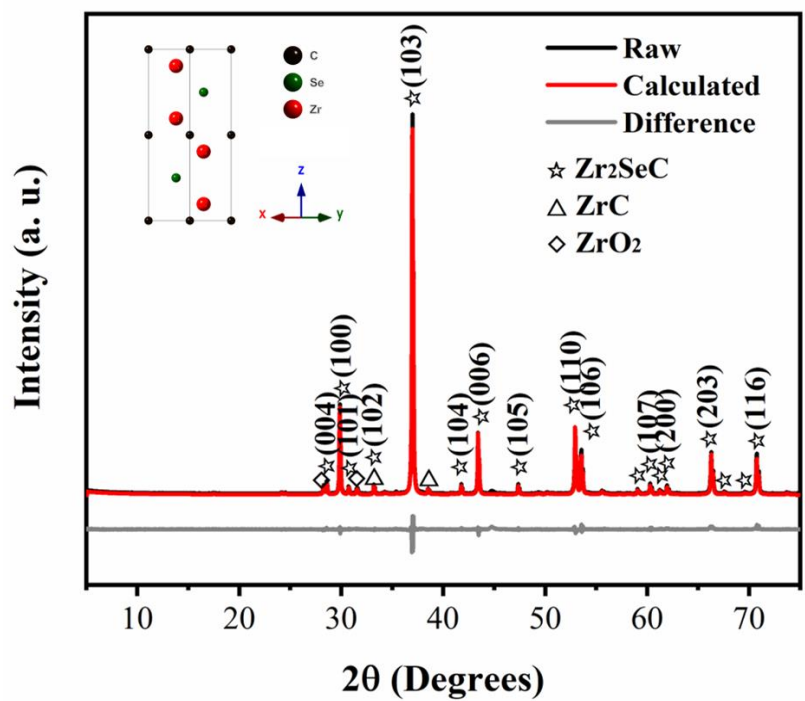

**Fig. 1.** Rietveld refinement of X-ray diffraction pattern of product from 3Zr-1.05ZrSe$_2$-1.95C mixed reactants after calcination at 1500 °C. The inset was simulated crystal structure of Zr$_2$SeC.

The as-synthesized products were characterized using X-ray diffraction in a full -2θ range peaks from 5° to 75° (**Fig. 1**). The Rietveld fitting results ($R_{wp}$=10.29%) indicated that the crystal structure belongs to hexagonal structure (P6$_3$/mmc, No. 194), which was the typical structure of MAX phase (the inset in **Fig. 1**). The respective coordinates of component elements were Zr (1/3, 2/3, 0.09628), Se (1/3, 2/3, 3/4), and C (0, 0, 0), and calculated lattice parameters were $a$=0.3462 nm and $c$=1.2518 nm, respectively. The TEM equipped with EDS was employed to analyze crystal structure (**Fig. S1**). The $d$-spacing of (10$\bar{1}$0) plane and (0002) plane were identified to be 0.3023 nm and 0.6278 nm, which help to obtain the lattice parameters of $a$ = 0.3491 nm and $c$ = 1.2556 nm. The first-principles calculations were also performed, and the results agreed well with above experimental results (**Table 1**): the calculated lattice parameters were $a$=0.3487 nm and $c$=1.2631 nm, and $z$ position in coordinate of Zr atom was 0.09629, verifying the formation of Zr$_2$SeC MAX phase. The crystal structure and precise atom arrangements within Zr$_2$SeC phase could be visually observed by advanced atom-resolved high-angle annular dark-field (HAADF) technique (**Fig. 2**). The color contrast of HAADF image directly relates to the atomic number (*Z*), i.e., the larger the *Z* value is, the brighter the atom will be. It could be easily distinguished that lines of bigger and brighter dots (Zr atoms, also marked in red for clearance) with zig-zag arrangement were mirrored by lines of smaller and dimmer dots (Se atoms, marked in light green), which is typical atomic arrangement of MAX phase viewing through zone axis [11$\bar{2}$0] (**Fig. 2a**). Meanwhile, the molar ratio of Zr to Se was semi-quantitatively estimated to be 2.07 that was much close to stoichiometry of 211 phase (**Table S1**).

**Table 1** Crystal parameters, density and atomic coordinates of Zr$_2$SeC based on XRD Rietveld method and first-principles calculations.

| | | a/nm | c/nm | ρ/g·cm$^{-3}$ | Zr | Se/S | C |
|---|---|---|---|---|---|---|---|
| Zr$_2$SeC | XRD refinement | 0.3462 | 1.2518 | 6.99 | (1/3, 2/3, 0.0963) | (1/3, 2/3, 3/4) | (0, 0, 0) |
| | Theoretical calculation | 0.3487 | 1.2631 | 6.83 | (1/3, 2/3, 0.0963) | (1/3, 2/3, 3/4) | (0, 0, 0) |
| | TEM analysis | 0.3491 | 1.2556 | | | | |
| Zr$_2$SC | XRD refinement | 0.3411 | 1.2147 | 6.14 | (1/3, 2/3, 0.1006) | (1/3, 2/3, 3/4) | (0, 0, 0) |
| | Theoretical calculation | 0.3438 | 1.2273 | 5.99 | (1/3, 2/3, 0.1007) | (1/3, 2/3, 3/4) | (0, 0, 0) |

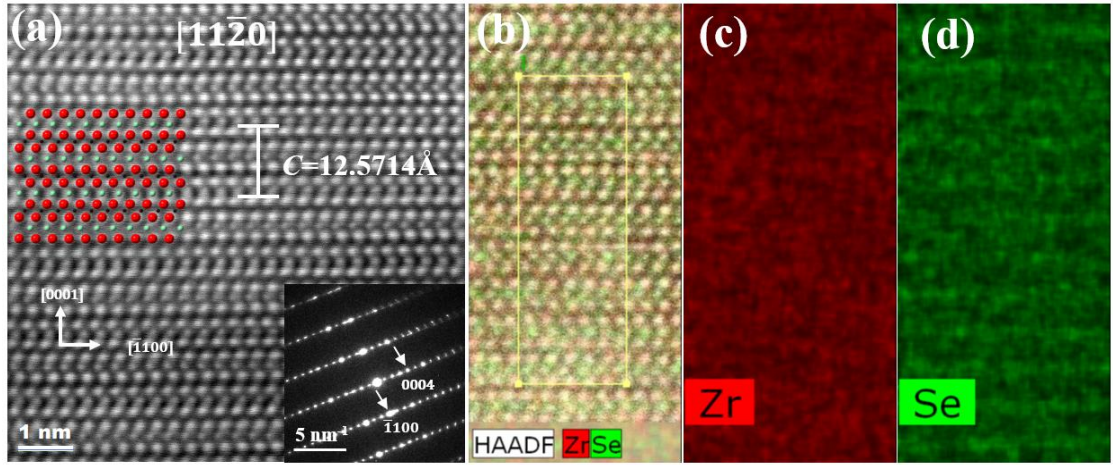

**Fig. 2.** Microstructure of $Zr_2SeC$ phase determined by TEM analysis: **(a)** Atom-resolved HAADF image, the insets were corresponding selected area electron diffraction (SAED). Colored dots are used to ease the discussion on atom arrangement. Atom-resolved EDS mapping images showing spatial distribution of **(b)** both Zr and Se, **(c)** Zr, and **(d)** Se elements.

A fully dense $Zr_2SeC$ bulk was *in situ* sintered in a Pulse-Electric-Current-Aided sintering furnace in order to explore its electrical and thermal properties. The relative density of 98.3% was finally achieved although trace impurities such as $ZrO_2$ and $ZrC$ particles always co-exist in final product (**Fig. 1**). The fracture surface clearly showed that there were few voids between/among $Zr_2SeC$ grains, ensuring the accurate interpretation of following electrical and thermal conduction behaviors (**Fig. S2**). Herein, a fully dense $Zr_2SC$ bulk was also sintered at 1600 °C for a comparative analysis (**Fig. S3**). The electrical conduction measurement on both $Zr_2SeC$ and $Zr_2SC$ bulks showed that these two MAX phases were conductors (**Table 2**). It was also reflected by the calculated energy band structures along the high-symmetry direction in the Brillouin zone (**Fig. S4**). The valence and conduction bands overlapped at Fermi levels, in both of $Zr_2SeC$ and $Zr_2SC$, implicate their metal-like conduction behaviors. The room-temperature electrical resistivity of $Zr_2SeC$ (1.57 μΩ·m) was a bit lower than that of $Zr_2SC$ (1.75 μΩ·m). This phenomenon could be explained by the density of states (DOS) in vicinity of the Fermi level ($E_F$). [22] The calculated total DOS of $Zr_2SeC$ was 1.48 states/eV/unit cell, which is a bit higher than that of $Zr_2SC$ (1.36 states/eV/unit cell). The partial DOS (PDOS) provided more details to understand the nature of chemical bonding (**Fig. 3b-c**). The most total DOS at $E_F$ was contributed by 4 $d$ electrons from Zr in $Zr_2SeC$ and $Zr_2SC$. Although the $s$ and $p$ electrons in Se and S were almost negligible in total DOS at $E_F$, they became significant in the range from –6.7 eV to –1.7 eV; which could be used to explain bond formation between Zr atoms and chalcogen atoms, *i.e.* the $p$ electrons of chalcogen atoms strongly interacted with $d$ electrons of Zr atoms. Moreover, the Zr-S bond was a bit stronger than Zr-Se bond, been confirmed by the charge-density difference images; that is, the accumulation of electrons around S atoms is much more intense in $Zr_2SC$ than that in $Zr_2SeC$ (**Fig. 3d-g**). On the other hand, the Ti 3$d$ is hybridized by the Al 3$p$ from –1.8 eV to –0.3 eV in $Ti_2AlC$, which is much weaker than above Zr-S and Zr-Se bonds. The strong bond strength is expected to benefit the phase stability and also the tuning of physical properties. [23]

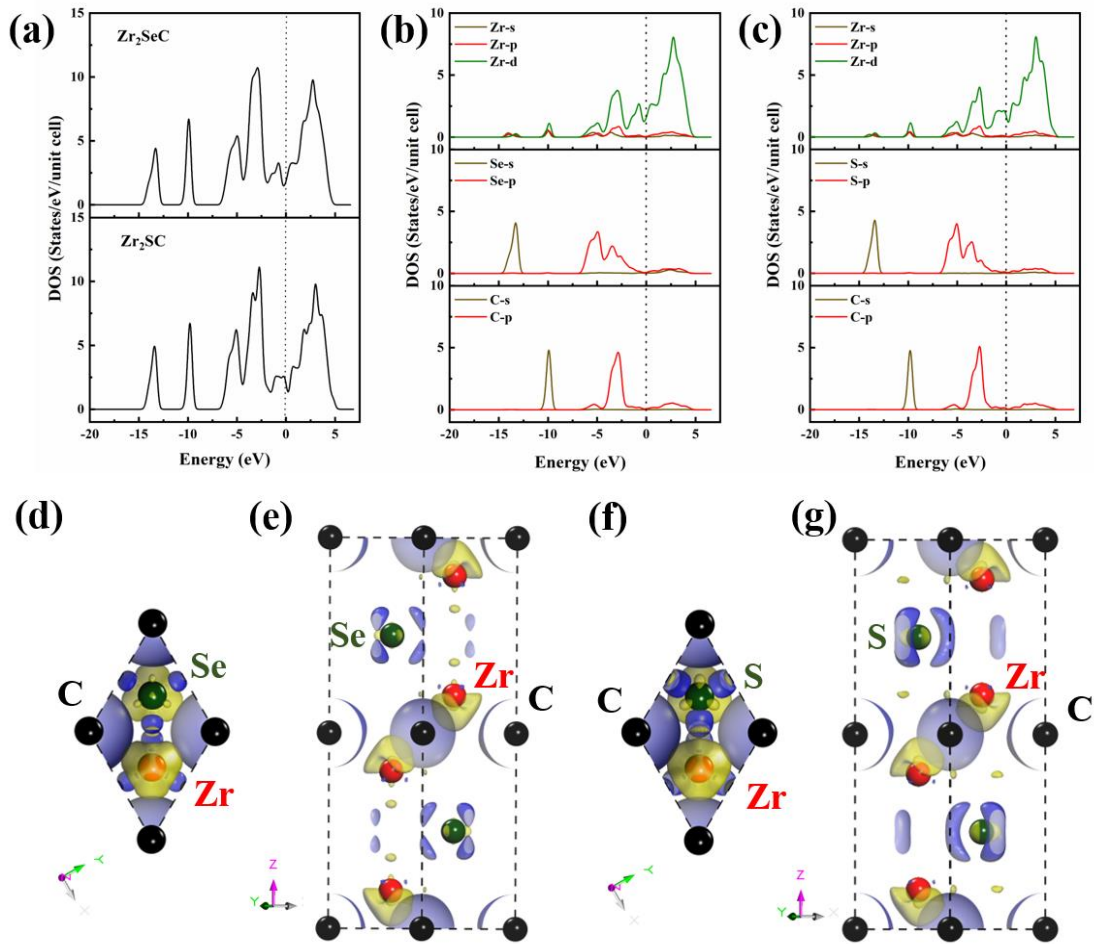

**Fig. 3.** Total density of states (DOS) **(a)** and partial density of states (PDOS) as a function of energy for **(b)** $Zr_2SeC$ and **(c)** $Zr_2SC$. Fermi level is aligned with 0 eV. Charge-density difference images of **(d-e)** $Zr_2SeC$ and **(f-g)** $Zr_2SC$ from different direction. The purple color shows the accumulation of electrons, and yellow color means the depletion of electrons.

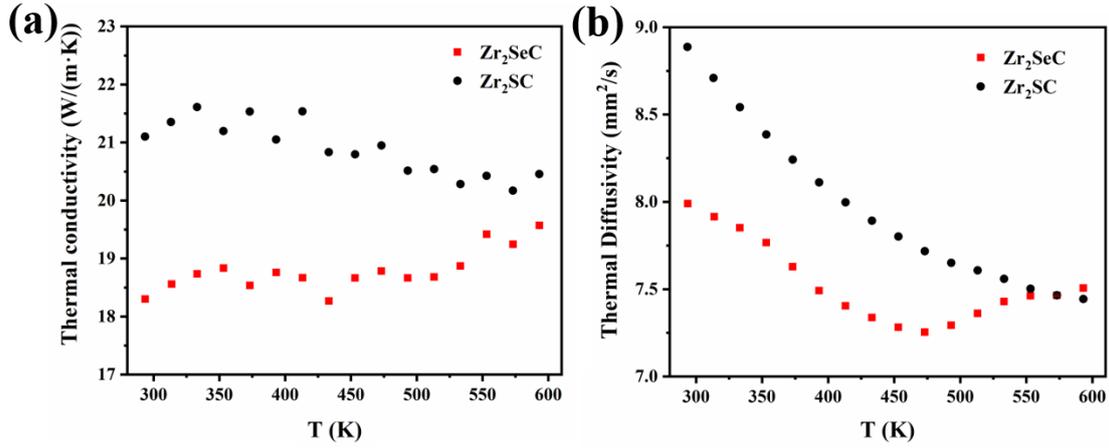

**Fig. 4. (a)** Thermal conductivity and **(b)** thermal diffusivity of $Zr_2SeC$ and $Zr_2SC$ along with temperature.

**Table 2** Electrical resistivity and thermal conductivity of $Zr_2SeC$ and $Zr_2SC$ at 293.15 K and 373.15 K.

| *T*/K | **Phase** | $\rho$/$\mu\Omega\cdot m$ | $\kappa$/W·m·K$^{-1}$ | $\kappa_e$/W·m·K$^{-1}$ | $\kappa_{ph}$/W·m·K$^{-1}$ |
|---|---|---|---|---|---|
| **293.15** | **$Zr_2SeC$** | 1.57 | 18.30 | 4.57 (25.0%) | 13.76 (75.0%) |
|  | **$Zr_2SC$** | 1.76 | 21.10 | 4.08 (19.3%) | 17.01 (80.7%) |
| **373.15** | **$Zr_2SeC$** | 1.67 | 18.54 | 5.47 (29.5%) | 13.02 (70.5%) |
|  | **$Zr_2SC$** | 2.01 | 21.53 | 4.55 (21.1%) | 16.98 (78.9%) |

The thermal conductivity ($\kappa$) is contributed by the charge carriers and phonon vibration, which could be evaluated by analyzing its temperature dependence (**Fig. 4a**). In this work, the $\kappa$ value of $Zr_2SeC$ and $Zr_2SC$ bulk were 18.30 W/m·K and 21.10

W/m·K at room temperature, respectively (**Table 2**). The thermal conductivity of Zr$_2$SC decreased when the temperature was increased from room temperature to 600 K (**Fig. 4a**), *i.e.* the typical thermal behavior dominated by phonon transport. In contrast, the thermal conductivity of Zr$_2$SeC kept unchanged around 18.5 W/m·K from room temperature up to 500 K and then slightly increased, indicating a different mechanism for thermal conductivity of Zr$_2$SeC. It is known that the electron-contributed thermal conductivity $\kappa_e$ can be deduced from the electrical resistivity through the Wiedemann-Franz law. [24]

$\kappa_e = L_0 T/\rho$

Where, $L_0$ is the classic Lorenz number (2.45×10$^{-8}$ W·Ω·K$^{-2}$). $T$ is the temperature (*K*), and $\rho$ is the electrical resistivity (*Ω·m*). Then the phonon contribution to the thermal conductivity ($\kappa_{ph}$) could be calculated as follows.

$\kappa_{ph} = \kappa - \kappa_e$

The $\kappa$, $\kappa_e$, and $\kappa_{ph}$ at room temperature and elevated temperature were calculated and shown in **Table 2**. Although the thermal conductivity of both phases was mainly dominated by phonon transport (75.2 % for Zr$_2$SeC, and 80.6% for Zr$_2$SC); the electronic contribution in Zr$_2$SeC is significantly higher than that in Zr$_2$SC, especially at elevated temperature. This discrepancy could be tentatively explained by the modification of M-A bond strength in these nanolaminated carbides. Stiff crystal structure benefits the phonon vibration as reflected a high phonon contribution to thermal conductivity in Zr$_2$SC. Selenium is larger in atomic size and has a smaller electronegativity than sulfur, thus the occupancy of Se in A site weakens the M-A bonding strength of Zr-S, as evident in an elongated bond length (2.791 Å for Zr-Se, and 2.701 Å for Zr-S, shown in **Table 3**) and an expanded *d*-spacing along *c* axis (**Table 1**). The weakened Zr-Se bond in Zr$_2$SeC would release more 4 *d* electrons near Fermi level in Zr atoms than that in Zr$_2$SC, and these delocalized electrons may be activated at elevated temperature. The bond length of Zr-Zr is slightly shorter in Zr$_2$SeC (3.156 Å) than that in Zr$_2$SC (3.170 Å), which also facilitates the electron transport of free 4 *d* electrons in plane of Zr atoms (**Table 3**). This phenomenon was evidently observed from the thermal diffusivity measurement. The thermal diffusivity of Zr$_2$SC decreased monotonically from room temperature to 600 K. However, the thermal diffusivity of $Zr_2SeC$ dropped a little before 475 K and then climbed up again to a value even higher than that of $Zr_2SC$. In order to fully understand the underlying physical mechanism of thermal conductance behavior of chalcogen-containing MAX phases, more work should be done in the future.

Table 3 The bond length in $Zr_2SeC$ and $Zr_2SC$ crystal structures

| MAX phases | M-A (Å) | M-X (Å) | M-M (Å) |
|---|---|---|---|
| $Zr_2SeC$ | 2.791 | 2.352 | 3.156 |
| $Zr_2SC$ | 2.701 | 2.338 | 3.170 |

## 4. Conclusion

In summary, a chalcogen-containing ternary MAX phase $Zr_2SeC$ was synthesized, and its crystal structure was comprehensively characterized in order to validate the occupancy of selenium element at A site of MAX phase. The first-principles calculations were also employed to simulate the ideal crystal structure of $Zr_2SeC$ and shed light on the chemical bonding as compared with $Zr_2SC$ counterpart. It was revealed that the electron contribution in thermal conductivity of $Zr_2SeC$ would be activated, and thus compensated the decline tendency at elevated temperature. These results reveal the importance of A-site elements on the tuning physical properties of MAX phases.

**Declaration of competing interest**

No potential conflict of interest was reported by the authors.

**Acknowledgements**

This study was financially supported by the National Natural Science Foundation of China (Grant No. 51902319, 51872302, 21805295), International Partnership Program of Chinese Academy of Sciences (Grant No. 2019VEB0008), Leading Innovative and

Entrepreneur Team Introduction Program of Zhejiang (Grant No. 2019R01003), Ningbo Top-talent Team Program, Ningbo Municipal Bureau of Science and Technology (Grant No. 2018A610005). Ke Chen acknowledges the support from the International Postdoctoral Exchange Fellowship Program of China (Grant No. YJ20180235).

**Supplementary information**

**MAX Phase Zr$_2$SeC and Its Thermal Conduction Behavior**

*Ke Chen [a,d,\*,#], Xiaojing Bai [b,#], Xulin Mu [c], Pengfei Yan [c], Nianxiang Qiu [a,d], Youbing Li [a,d], Jie Zhou [a], Yujie Song [a,d], Yiming Zhang [a,d], Shiyu Du [a,d], Zhifang Chai [a,d], Qing Huang [a,d,\*]*

[a] Engineering Laboratory of Advanced Energy Materials, Ningbo Institute of Materials Technology and Engineering, Chinese Academy of Sciences, Ningbo 315201, China.

[b] School of Materials Science and Engineering, Anyang Institute of Technology, Anyang 455000, China.

[c] Beijing Key Laboratory of Microstructure and Properties of Solids, Beijing University of Technology, Beijing 100124, China.

[d] Qianwan Institute of CNiTECH, Ningbo 315336, China.

[#] These authors contributed equally to this work.

\* Corresponding authors.

E-mail addresses: chenke@nimte.ac.cn (Ke Chen), huangqing@nimte.ac.cn (Qing Huang).

**Figures and Table**

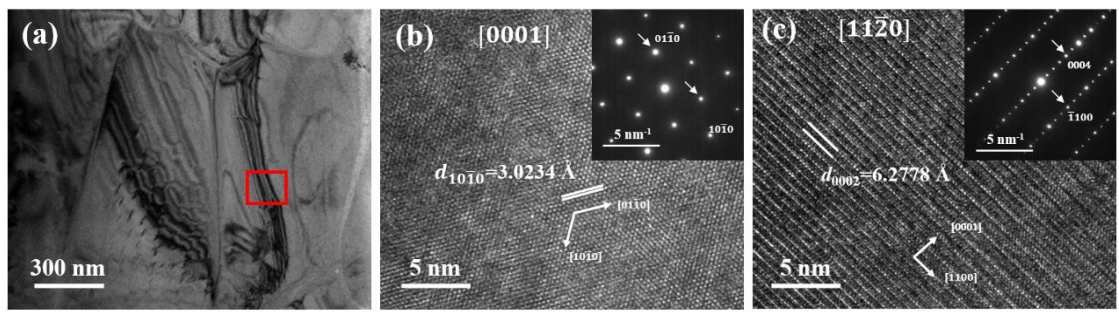

**Fig. S1. (a)** TEM and High-resolution (HR)-STEM image of $Zr_2SeC$ along the **(b)** [0001] and **(c)** [11$\bar{2}$0] direction. The insets were the corresponding selected area electron diffraction (SAED).

Table S1 Elemental analysis of the selected area in **Fig. S1a**

| Element | Zr | Se | C | O | Total |
|---|---|---|---|---|---|
| at. % | 56.07 | 27.09 | 10.60 | 6.25 | 99.99 |

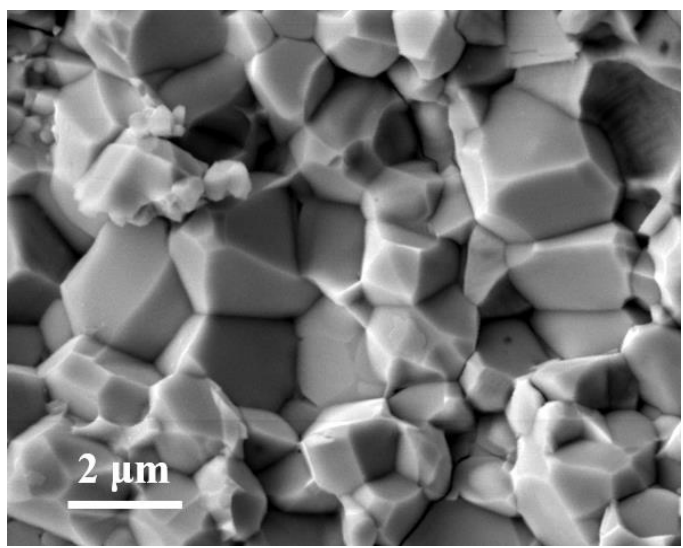

**Fig. S2.** Fracture morphology of $Zr_2SeC$ bulk.

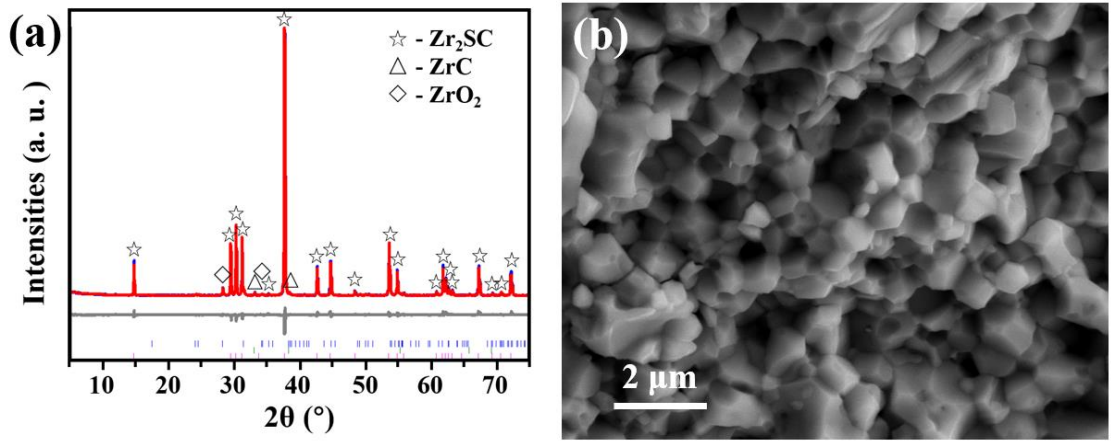

**Fig. S3. (a)** Rietveld refinement of X-ray diffraction pattern and **(b)** fracture morphology of sintered $Zr_2SC$ bulk.

The main phase of the as-sintered bulk was $Zr_2SC$, with minor impurity of $ZrO_2$ (JCPDS # 78-0047) and ZrC (JCPDS # 89-4054). The refined lattice parameters of $Zr_2SeC$ phase, $a$ and $c$, were 0.3411 nm and 1.2147 nm, respectively ($R_{wp}$=11.31%). The relate density determined by Archimedes method was 99.2%. The fracture surface showed that there were few voids in the bulk.

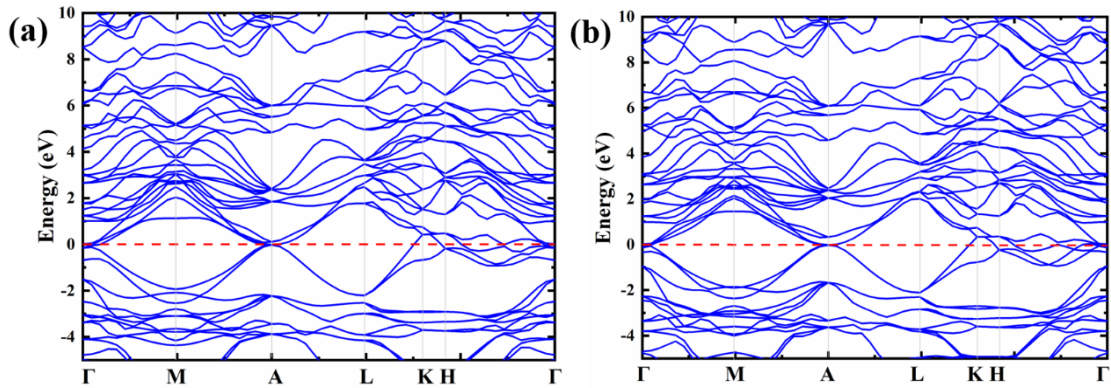

**Fig. S4.** Band structure along the principal high-symmetry directions in the Brillouin zone of **(a)** $Zr_2SC$ and **(b)** $Zr_2SeC$.